\documentclass[11pt,a4paper,amsmath,nofootinbib]{article}

\usepackage{jheppub}

\usepackage{epsfig}

\usepackage{graphicx}

\title{A Rationale for Long-lived Quarks and Leptons at the LHC:
Low Energy Flavour Theory} 

\author[a]{O. J. P.~\'Eboli,}
\author[b]{C. A.~Savoy}
\author[a]{and R.~Zukanovich Funchal}

\affiliation[a]{Instituto de F\'{\i}sica, Universidade de S\~ao Paulo, \\ 
 C.\ P.\ 66.318, 05315-970 S\~ao Paulo, Brazil}
\affiliation[b]{Institut de Physique Th\'{e}orique, CEA-Saclay\\
Orme des Merisiers, 91191 Gif-sur-Yvette, France}

\emailAdd{eboli@fma.if.usp.br} 
\emailAdd{carlos.savoy@cea.fr}
\emailAdd{zukanov@if.usp.br}

\abstract{In the framework of gauged flavour symmetries, new fermions
  in parity symmetric representations of the standard model are
  generically needed for the compensation of mixed anomalies. The key
  point is that their masses are also protected by flavour symmetries
  and some of them are expected to lie way below the flavour symmetry
  breaking scale(s), which has to occur many orders of magnitude above
  the electroweak scale to be compatible with the available data from
  flavour changing neutral currents and CP violation experiments. We
  argue that, actually, some of these fermions would plausibly get
  masses within the LHC range. If they are taken to be heavy quarks
  and leptons, in (bi)-fundamental representations of the standard
  model symmetries, their mixings with the light ones are strongly
  constrained to be very small by electroweak precision data. The
  alternative chosen here is to exactly forbid such mixings by
  breaking of flavour symmetries into an exact discrete symmetry, the
  so--called proton--hexality, primarily suggested to avoid proton
  decay. As a consequence of the large value needed for the flavour
  breaking scale, those heavy particles are long--lived and rather
  appropriate for the current and future searches at the LHC for
  quasi--stable hadrons and leptons.  In fact, the LHC experiments
  have already started to look for them.}
 

\arxivnumber{}

\begin{document}
\maketitle

\section{Introduction}

Experiments at the Large Hadron Collider (LHC) have published their
first results on the production of quasi--stable hadrons that are
stopped and decay in the detectors~\cite{exp,exp2}.  Besides special
cases of supersymmetric and extra-dimensional models, other models of
different sorts also predict heavy particles with delayed
decays~\cite{Fairbairn:2006gg}, and many will be checked in turn at
the LHC. In this letter, we discuss a new class of metastable
particles suitable for those searches that arise in the framework of
flavour models.

We argue that, in spite of the generic bound on the symmetry breaking
scales of gauge flavour theories from rare
transitions~\cite{ArkaniHamed:1999yy} being many orders of magnitude
above the LHC reach, heavy quarks and/or leptons are plausibly
expected with masses in the TeV range. Indeed, their masses can be
much below the flavour breaking scale because they would be as
protected by the flavour symmetries as the lighter Standard Model (SM)
quarks that get masses much below the Fermi scale. As elaborated
below, this may happen if the \emph{raison d'\^etre} of those heavy
fermions is to cancel the anomalous couplings between the flavour and
the SM gauge bosons induced by the light fermion sector.

The heavy fermion metastability is mainly due to an exact discrete
symmetry that survives the flavour and electroweak symmetry
breakings. This additional symmetry is needed to prevent proton decay
without forbidding neutrino masses. In order to build a consistent
effective theory, we then assume a conserved symmetry, the so--called
proton hexality~\cite{Dreiner:2007vp}, which can be written as
$(B-3L)/6$ in terms of the usual baryon and lepton numbers. Since we
assume that it results from the breaking of the continuous flavour
symmetry, it is local and non--anomalous\footnote{In the
  supersymmetric version~\cite{Savoy:2010sj}, it replaces the
  $R$-parity.  We refer to this paper for a more detailed discussion
  of some issues below}.

Because these new fermions can be consistently endowed with
non-standard discrete baryon/lepton numbers they do not mix at all
with SM fermions but can decay into three light fermions through
dimension--six operators. Therefore their decay lengths
are\footnote{The symbol ``$\sim$'' is used here to denote equality up
  to an $O(1)$ factor.} $ c\, \tau_F \sim {48\pi ^3 \Lambda^4}
m_F^{-5} $, where their masses are $m_F = O(\mathrm{TeV})$ and
$\Lambda^{-2}$ is the coefficient of the corresponding four-fermion
interactions. From the analysis of the analogous flavour changing
neutral current (FCNC) dimension--six operators for light fermions,
flavour physics puts a limit $\Lambda > 5\cdot10^4\,
\mathrm{TeV}$~\cite{Isidori:2010kg}. Hence $ c\tau_F > 2 \,
\mathrm{km}$ - actually, much more in the explicit models displayed
below.

Of prime importance is the fact that the new states belong to real
representations of the SM gauge symmetry, so that their masses mostly
arise from their couplings to the flavon(s) rather than to the Higgs boson, 
and mass mixings with light fermions are forbidden by the discrete symmetry. 
Thus the severe bounds on those mixings are naturally avoided and their 
contributions to electroweak precision tests (EWPT) are quite suppressed.
\\

\section{A model}

\noindent For the sake of argument, we take the instance of a single
charge abelian flavour symmetry group $U(1)_X$ broken by the vacuum
expectation value (\emph{v.e.v.}) of a single complex scalar ($\phi$),
the so-called flavon, and assign (different) $X$-charges to the quarks
and leptons of either chirality in the three families so to yield an
acceptable description of the observed mass hierarchies and
mixings. Furthermore, by a suitable combination with the hypercharge,
we choose for the Higgs, $X(H)=0$ and normalize it so that $X(\phi)=
-1$. Most important in our analysis are the chiral charge differences
of quarks and leptons, $\chi_f^{ij} = X(f^i_L) - X(f^j_R) \equiv f^i_L
- f^j_R$, where $f_L= q, l $, $f^i_R = u, d, e$, and $i,\,j =1,2,3$
are family indices. All the SM model indices are dropped, in a
notation that is quite standard for the SM fermions of both
chiralities. The chiral charge differences forbid the couplings of all
fermions but the top quark to the Higgs at the renormalizable level,
providing the needed chiral protection for their masses.

This defines a Froggat-Nielsen effective Lagrangian~
\cite{Froggatt:1978nt,Bijnens:1987ff} with a cutoff $\Lambda$ way
above the electroweak scale after integrating out the flavon and the
$U(1)_X$ gauge boson, denoted $X_{\mu}$. We then basically follow the
steps in the supersymmetric version in~\cite{Savoy:2010sj}, to which
we refer for more details, although the presence of scalars there
makes the effective theory, as much as its phenomenology, very
different from the one discussed below (in particular, the
supersymmetric heavy states are short-lived).
\\

\section{Charged fermion and neutrino masses}

\noindent The flavon \emph{v.e.v.}, $\langle \phi \rangle = \epsilon
\Lambda$, breaks the flavour symmetry and allows for the fermions to
acquire masses from higher dimension operators with a Higgs field and
$\chi_f^{ij}$ insertions of the field $\phi$ ($\phi^{\dag}$) and
divided by $\chi_f^{ij}$ factors of the cutoff $\Lambda$. The mass
matrices are then $ m^{ij}_f \sim \epsilon^{\vert \chi^{ij}_f \vert}
\, v$, where $ v=174 \, {\rm GeV}$ is the \emph{v.e.v.} of the Higgs
field and $ f= u,\, d,\, l$. From the masses and (left-handed) mixings
one can find several solutions for the $X$-chiralities matrices
$\vert\chi_u^{ij}\vert,$ $\vert\chi_d^{ij}\vert$ and
$\vert\chi_l^{ij}\vert$ that can be found in the
literature~\cite{Ibanez:1994ig, Binetruy:1994ru, Dudas:1995yu,
  Dudas:1995eq}. Here we mostly need their traces: Tr$\vert\chi_f\vert
\approx \mathrm{ln \, det}(m_F/v) / \mathrm{ln}\epsilon$ and the value
of the parameter $\epsilon$ that comes out close to the Cabibbo angle,
$\epsilon \sim \theta_C \approx 0.2$. Thus one finds: Tr$\vert \chi_u
\vert \approx 12$, Tr$\vert \chi_d\vert \approx \mathrm{Tr}\vert
\chi_l \vert \approx 15$. The fits are better if Tr$\chi_u$ and
Tr$\chi_d$ have the same sign, which we choose to be positive.

The effective dimension--five operator $(\nu_i \cdot H)(\nu_j \cdot H)
/ \Lambda$ has charge $X = l_i +l_j$, therefore, the resulting
neutrino mass matrix becomes $m_{\nu}^{ij} \sim \epsilon^{\vert l_i
  +l_j \vert} \, v^2/ \Lambda$. Experiments allow for only a little
hierarchy in the eigenvalues and relatively large mixings, so the best
one can do in our case is to choose the same charge $l_i = l$ for all
families, so that,
\begin{equation}
  \epsilon^{\vert 2 l\vert} \sim \frac{\sqrt{\Delta m^2_{\rm atm}} \, 
   \Lambda}{v^2} = \frac{\Lambda}{6 \times 10^{11} \, \rm TeV}\, . 
\label{eq:epsell}
\end{equation}
This fixes ${\vert l \vert}$ for a given cutoff. By imposing $\Lambda
\gtrsim 5\cdot10^{4} \, \rm TeV$ to meet the requirements from flavour
physics discussed above, it follows that ${\vert 2 \, l \vert} \le
11$.
\\

\section{Exact discrete symmetry}
\label{sec:esim}

\noindent Omitting all Lorentz and SM indices, there are six $B$ and
$L$ violating four--fermion operators that can be schematically
written as: $lqqq$ (2 operators), $eudu$ (2 operators), $euqq$, and
$lqdu$~\cite{Weinberg:1979sa, wilzee}. They are related to proton
decay and preserve $B - L$. Thus, they are forbidden by the discrete
symmetry generated by the charge $Z'= B/6 - L/2 \ \mathrm{mod}\, 1$,
defining proton hexality, which stabilizes the proton. One can also
combine $Z = (Z' - Y/3) \ \mathrm{mod}\, 1$ which defines a ${Z}_6$
symmetry as can be seen from the $Z'$ and $Z$ quantum numbers
displayed in table~\ref{tab1}. Neutrino Majorana masses are allowed
and the flavon $\phi$ is invariant. We embed this symmetry in the
flavour one $U(1)_X$, so that it is local and non-anomalous in the
most economical way. With this choice, the $U(1)_X$ symmetry is broken
by the $\phi$ \emph{v.e.v.} into the discrete one and $Z'$ coincides
with the fractional part of X.  Notice that $Z'$ is vector-like to
allow for fermion masses, family-independent to allow for family
mixings in weak interactions, and anomaly-free with respect to the
SM. Since $B-L$ allows for proton decay, one is left with $Z'$ to avoid
a fast proton decay.

Concerning the new heavy states to be introduced later on, one must
efficiently protect the light fermion mass matrices and efficiently
suppress the mixing with the light ones by either giving exotic
quantum numbers with respect to the SM gauge group, or by assigning
different $Z'$ charges to them. In the first case the mixings will be
reduced by factors of $v/\Lambda$, but the structure in the mass
matrices might be affected in a model-dependent way, so that we choose
the second one where the mixings are just forbidden. It seems natural
to compensate the anomalies from the SM sector by a choice of heavy
fermions with, instead, the same SM quantum numbers, and denote them
by capitals as $F= Q,\, L,\, U,\, D,\, E$, with the same chiralities
as the light ones ($Q\equiv Q_L ,\, U\equiv U_R$, etc.). Their parity
conjugated states are denoted $F'$. Furthermore, the notation for the
charges is simplified below: $X_F \equiv F$.  \\

\begin{table}
\begin{tabular}{|c|c|c|c|c|c|}
\hline 
 $f$ & $SU(3)$& $SU(2)$ & $Y$ & $Z_6$ & $18 Z'$ \\
\hline 
 $q_{(L)}$ & $3$ & $2$ & $1/6$ & $ 0$ &  1 \\
\hline
 $u_{(R)}$ & $3$ & $1$ & $2/3$ & $-1/6$ &  1 \\
 \hline
 $d_{(R)}$ & $ 3$ & $1$ & $-1/3$ & $1/6$ &  1 \\
\hline
 $l_{(L)}$ & $1$ & $2$ & $-1/2$ & $ -1/3$ &  -9 \\
\hline
 $e_{(R)}$ & $1$ & $1$ & $-1$ & $-1/6$ &  -9 \\
\hline 
\end{tabular}
\begin{tabular}{|c|c|c|c|c|c|c|}
\hline 
 $F$ & $F'$ & $SU(3)$ & $SU(2)$ & $Y$ & $Z_6$ & $18 Z'$ \\
\hline 
 $Q_{(L)}$ & $Q'_{(R)}$ &  $3$ & $2$ & $1/6$ & $1/3$ &  7 \\
\hline
 $U_{(R)}$ & $U'_{(L)}$ & $ 3$ & $1$ & $2/3$ & $ 1/6$ &  7 \\
 \hline
 $D_{(R)}$ & $ D'_{(L)}$ & $ 3$ & $1$ & $-1/3$ & $ 1/2$ &  7 \\
\hline
 $L_{(L)}$ & $L'_{(R)} $ & $1$ & $2$ & $-1/2$ & $0$ &  -3 \\
\hline
 $E_{(R)}$ &  $E'_{(L)}$ & $1$ & $1$ & $-1$ & $1/6$ &  -3 \\
\hline 
\end{tabular}
\caption{Notation and quantum numbers of light and heavy fermions with
$Z= Z' - \frac{1}{3} Y$. In the text the chiralities $(L,R)$ are omitted. }
\label{tab1}
\end{table}

\section{Heavy quarks and leptons}

\noindent There are arbitrarinesses in the choice of heavy states
needed for the cancellation of anomalies. Let us assume asymptotic 
freedom so that the total number of heavy quarks is limited to ten. If we 
further assume whole families of heavy fermions analogous to the SM 
ones together with their parity conjugated states, there can be at most 
two of them. For simplicity, here we concentrate on just one heavy family, 
the generalization to two being straightforward as much as for the case of 
incomplete families.

The main contributions  to the masses of the heavy fermions, $m_F$, are 
reduced from their natural scale $\Lambda$ by their X-chiralities, defined 
as $\chi_F = F_L - F_R $ so that,
\begin{equation}
m_F \sim \epsilon^{\vert \chi_F \vert} \, \Lambda \sim (3 \times 10^{12})\,
\epsilon^{\vert \chi_F \vert + \vert 2 l \vert}  \,v
\label{eq:m_F}
\end{equation}
where (\ref{eq:epsell}) has been used. 

In the analysis of the model that follows, there are two bounds one
can derive on $m_F$ which imply on limits on the largest $\vert \chi_F
\vert$. A lower bound that all new fermions have to oblige since they
have not been observed yet, i.e., $(m_F)_{\rm min} \gtrsim v$,
and another that selects models with fermions that can be reasonably
produced at the LHC, by an upper bound $(m_F)_{\rm min} \lesssim
\epsilon^{-2}v \sim 4 \, {\rm TeV}$ for the lightest heavy fermion
. Therefore we impose the condition:
\begin{equation}
16 - 2 \vert l \vert \le {\vert \chi_F \vert}_{\rm max} \le 18 - 2 
\vert l \vert \label{eq:chiFlmax} .
\end{equation}
\noindent Notice that $ {\vert \chi_F \vert}_{\rm max}$ corresponds to
the lightest long-lived particle (LLLP), although more than one state
could share the same value of $\vert \chi_F \vert$ (see one example
below), or the next lightest particle could be more easily
produced. Finally, one can introduce two whole or incomplete families
to compensate the anomalies. Yet, some states could be much lighter
than the others, leading to results similar to the one family case in
analogy with what happens with the three light families.
\\



Now, since the new states cannot be stable for cosmological reasons,
one must allow for their coupling to three light fermions via
dimension--six operators. Their SM quantum numbers allow only for the
analogous of the proton--decay four--fermion operators described
above, see Sec.~\ref{sec:esim}, with any one light state replaced by
its corresponding heavy one. This leads to the following patterns of
decays:
\begin{eqnarray}
& L \rightarrow \bar{q}\bar{q}\bar{q}\qquad L \rightarrow
  \bar{u}\bar{d}\bar{q}\qquad E \rightarrow
  \bar{u}\bar{q}\bar{q}\qquad E \rightarrow \bar{u}\bar{u}\bar{d}
  \nonumber \\ & Q \rightarrow \bar{q}\bar{q}\bar{l}\qquad Q
  \rightarrow \bar{d}\bar{u}\bar{l}\qquad Q \rightarrow
  \bar{q}\bar{u}\bar{e}\qquad D \rightarrow \bar{q}\bar{u}\bar{l}
  \nonumber \\ & \ \ \ \ \ \ D\rightarrow \bar{u}\bar{u}\bar{e}\qquad
  U \rightarrow \bar{q}\bar{u}\bar{l}\qquad U\rightarrow
  \bar{q}\bar{q}\bar{e}\qquad U \rightarrow
  \bar{d}\bar{u}\bar{e}\qquad
\label{eq:decays}
\end{eqnarray}
where all indices have been skipped.

The heavy fermion $Z'$ are to be chosen \emph{ad hoc}, and their
parity conjugated states have the same values to allow for singlet
masses. They are displayed in table~\ref{tab1}: heavy quarks have $Z'
= 7/18$, heavy leptons have $Z' = - 1/6$ like antinucleons.

Of most relevance are the flavour charges since they control the
suppression of the decay rates: they select the main modes in terms of
light fermion flavours.  In most of the cases, lifetimes are
increased by many orders of magnitude with respect to the lower bound
of a few meters (cf., the examples below).  The experimental long-life
signatures are not very sensitive to the lifetimes as far the decay is
always delayed enough, but the family dependence of the decay products
are model dependent.  \\

\section{Electroweak precision tests}

\noindent Besides the dominant masses above, the heavy states get mass
contributions also from their couplings to the Higgs, analogous to the
SM ones. The set of electroweak precision tests (EWPT) are very
sensitive to these couplings, being the most important new physics
contributions summarized by the oblique
parameters~\cite{Peskin:1991sw}.  Let us concentrate, for
definiteness, into the system $\{ Q, U, Q', U' \}$, keeping in mind
the definition of their $X$-chiralities. The order of magnitude of
mass terms are given by:
\begin{equation}
\epsilon^{\vert \chi_Q \vert}\Lambda \bar{Q'}Q + \epsilon^{\vert
  \chi_U \vert}\Lambda \bar{U}U' +\epsilon^{\vert \chi_{QU} \vert}v
\bar{U}Q +\epsilon^{\vert \chi_{Q'U'} \vert}v \bar{Q'}U' +
\mathrm{h.c.}
\label{eq:heavymass}
\end{equation}
\noindent where $\chi_{QU} = Q-U$ and $\chi_{Q'U'} = Q' - U'$. The
last two (`non-diagonal') masses are reduced by a factor $v/\Lambda <
\epsilon^8$ with respect to the dominant (`diagonal') ones, however,
their actual ratios depend on model dependent flavour factors. The
other mass terms are analogously obtained by the replacements $Q,U
\rightarrow Q,D$ and $Q,U \rightarrow L,E$.

The heavy fermion loops contribute to the S and T parameters with,
respectively, two and four insertions of Higgs in the new fermion
lines, given by the appropriate combinations of the non-diagonal
masses $m_{QU}$, $m_{QD}$, $m_{LE}$, $m_{Q'U'}$, $m_{Q'D'}$,
$m_{L'E'}$. Roughly, S and T are proportional to products of two mass
ratios , e.g., $m_{QU} m_{Q'U'}/m_{Q}^2\sim \epsilon^{\vert \chi_{QU}
  \vert +\vert \chi_{QU} \vert-2\vert \chi_Q \vert}(v/\Lambda)$, and
similar ones. General upper bounds can be obtained from the
approximations displayed in ref.~\cite{Martin:2009bg} for the
contributions $\Delta S$ and $\Delta T$ from the heavy quark
sector. Replacing all the non-diagonal masses in the numerators of
expressions therein by the largest one, denoted $\hat{m}$, and the
diagonal ones in the denominators by the smallest one, $\hat{M}$,
leads to the upper limits:
\begin{equation}
 \Delta T \lesssim \frac{N_c}{2\pi} \,\frac{\hat{m}^4}
 {v^2 \hat{M}^2} \qquad
 \Delta S \lesssim \frac{N_c}{8\pi} \,\frac{\hat{m}^2}
 {\hat{M}^2} 
\label{eq:ewpt}
\end{equation}
where $N_c =3$ for heavy quarks and $1$ for heavy leptons. Since
$\hat{m} \lesssim v$ while $\hat{M} \gtrsim v$ from the existent
bounds, the contributions of the heavy fermions might be dangerous
only when\footnote{We are assuming only one non--diagonal mass to be
  dominant. As asserted in ref.~\cite{{Martin:2009bg}},
  (\ref{eq:ewpt}) are overestimates for $\hat{m} \sim \hat{M}$.}
$\hat{m} / \hat{M}$ is $O(1)$, namely, when both are $O(v)$.
Therefore, the model would be constrained only in a very small portion
of its parameter space.  Indeed, the discrete symmetry introduced to
protect the mass matrix textures also plays a role for the protection
of the EWPT parameters by avoiding additional mixed couplings to the Higgs.  \\

\section{FCNC and CP violations}

\noindent The main obstacle to lower the flavour symmetry breaking
scale comes from the impressive description of flavour changing and CP
violating processes by the SM, leaving in some cases almost no space
for new physics contributions. In our approach here, the dangerous
operators are $B=L=0$ four--fermion operators with arbitrary flavour
structure, suppressed by $\Lambda^2$, as well as powers of $\epsilon$
due to $X$-charge imbalance, which are relatively model
dependent. Other dimension--six operators, $f_R^* H f_L \mathcal{F}$
where $\mathcal{F}$ stands for the photon or the gluon field strength,
give rise to flavour changing magnetic moments and electric dipole
moments, hence strong limits on $\Lambda$.

However the utmost limit on the cutoff $\Lambda$ turns out to be
generic in our framework~ \cite{ArkaniHamed:1999yy}. Indeed, the
exchange of massive flavour gauge bosons give rise to current-current
interactions of range $\epsilon \Lambda$, which of course are flavour
diagonal in the basis where $X$ is diagonal. Since the fermion charges
are different, in going to the mass eigenstate basis, the mixings in
flavour currents are given by the mass matrices. Thus for the $ s
\rightarrow d$ transitions one expects a mixing $O(\theta_C) \sim
\epsilon$. Comparison with rare K-physics
processes~\cite{Isidori:2010kg} requires $\Lambda > 5\cdot 10^4\,
\mathrm{TeV}$, as already asserted, puts the lower bound on the cutoff.  \\

\section{Anomaly cancellations}

\noindent Let us now turn to the main ingredient, the cancellation of
anomalies introduced by the SM fermion loops. Each one has to be
canceled by the heavy sector comprising fermions with SM quantum
numbers similar to the light ones. For simplicity we write the
expressions for only one regular family of heavy fermions, the
generalizations being evident. The anomalies with one flavour boson
coupling to the parity conserving gauge bosons of QCD and QED, depend
only on the X-chiralities of the fermion in the loops. They can be
combined\footnote{See, e.g., \cite{Savoy:2010sj} for details.}
into two relations:
\begin{eqnarray}
 2\chi_Q + \chi_U + \chi_D & = & - \mathrm{Tr}(\chi_u + \chi_d) \quad  
 (\, \approx - 27\,) ,\label{eq:anomalyS} \\
\chi_L + \chi_E - \chi_Q - \chi_D & = & -\mathrm{Tr}(\chi_l - \chi_d)\quad
 (\ge -30,\ \le 0)      ) ,  
\label{eq:anomalyQ+S}
\end{eqnarray}
\noindent where the values evaluated above are used to obtain the
value and limits in brackets, since the signs of $\chi_l$ are not fixed.

The cancellation of the anomalous coupling of $X_{\mu}$ to two $W'$s 
results in:
\begin{equation}
3 \chi_Q + \chi_L = -\mathrm{Tr}(3q + l) \approx -3\mathrm{Tr}(q) + 3l\,. 
\label{eq:anomalyWW}
\end{equation}
\noindent depends on the sum of $X$-charges that are very
model-dependent. Notice that the contribution of the fractionary
charges is $9Z'_q - 3Z'_l$, which, of course, vanishes for $B-L$ and
is integer ($-1$) for $Z' = B/6 - L/2$ so that the heavy states can
cancel the anomaly. Finally, the anomalous coupling of the photon to
two $X_{\mu}$'s is zero if:
\begin{equation}
\sum_{F=Q,U,D,L,E)}\mathrm{Tr}Q_F  \chi_F \, (F_L + F_R ) =\  - 
\sum_{f=u,d,l} \mathrm{Tr} Q_f \chi_f \, (f_L + f_R )
\label{eq:anomalyQXX}
\end{equation}
where $Q_f, Q_F$, are the electric charges and $\mathrm{Tr}$ is taken 
on all SM and flavour indices. This equation also depends explicitly 
on the $X$-charges.

Clearly, since there are many more parameters than conditions, there
are many solutions to these equations even with a single heavy family.
However, the values taken by \emph{r.h.s.}'s in (\ref{eq:anomalyS})
and (\ref{eq:anomalyQ+S}) indicate that for many solutions the LLLP would
fit into the visibility range (\ref{eq:chiFlmax}).

Interestingly enough, (\ref{eq:anomalyS}) implies that ${\vert
  \chi_F\vert}_{\rm max} \ge 7$, hence $\vert 2l \vert \le 11$. From
(\ref{eq:epsell}), this is equivalent to the limit on the cutoff from FCNC
experiments, i.e., in the models discussed here with only one
heavy family, the solutions to the anomaly conditions with $\Lambda <
5\cdot10^5\, \mathrm{TeV}$ would require a metastable fermion too
light to have escaped observation.  \\

\section{Case studies}

\noindent From (\ref{eq:anomalyWW}), one sees that the solutions
depend quite strongly upon Tr$q$ and, of course, upon $l$ which also
fix the scale $\Lambda$, and Tr$\chi_l$ which is not fixed by the
charged lepton masses.  For the sake of illustration, we consider only
a few simple choices that lead to satisfactory mass matrices. Let us
first introduce a notation for the integer part of the $X$-charges:
$x_f = X_f - Z'_f$. The results are shown in table~\ref{tab3}. Of
course there are many more solutions.

Notice that only relatively high scales, $\Lambda \sim
5\cdot10^5\,(10^7)$ TeV for $x_l = 5$ or $-4 \ (4)$ are obtained in
these examples. These translate into very long lifetimes.  In one case
an incomplete family is enough to compensate the anomalies and the
metastable lepton is quite light.  Heavy quarks decay with a very
energetic lepton in the final state, with a characteristic energy
distribution. \\

\begin{table}[t]
\begin{tabular}{|c|c|c|c|c|}
\hline  $x_{Q}$\quad $x_{U}$\quad $x_{D}$\quad $x_{L}$\quad $x_{E}$& 
\textbf{LLLP} & \textbf{Mass} & \textbf{Decay Modes} 
&  $\mathbf{c\, \tau}$\\
$x_{Q'}$~~$x_{U'}$\quad$x_{D'}$~~$x_{L'}$\quad$x_{E'}$ & & & & 
\\ \hline
\multicolumn{5}{|c|}{$\chi_l = 15$\quad$x_l=5$\quad$x_d=(3 \ 3 \ 3)$\quad 
$x_q=x_u=(4 \ 2 \ 0)$\quad$x_e=(2\   0 -2)$} \\ 
\hline 
$-8$\quad$+3$\quad$+6$\quad$-8$\quad$+5$ & $Q$ or $L$ & 
$m_Q \sim m_L \sim \epsilon^8 \,\Lambda$& $Q \to ldu , que$ & 
$\sim 4 \times 10^{3}$ km \\ 
$0$\quad$-3$\quad$+1$\quad~$\ \,0$\quad\, $\ \,0$& & $ \sim 1$ TeV & $L 
\to duq $ & $\sim2 \times 10^{4}$ km\\ 
\hline \hline  
$-6$\quad$+1$\quad$+2$\quad$-5$\quad$+5$ & $Q$ or $L$ 
& $m_Q \sim m_L \sim  \epsilon^8 \,\Lambda$& 
$Q \to que, ldu$ & $ \sim  4 \times 10^{3}$ km \\ 
$+2$\quad$-4$\quad$-4$\quad$+3$\quad$-1$ & & $ \sim 
1$ TeV & $L \to duq $ & $\sim 4 \times 10^{3}$ km\\   
\hline
\multicolumn{5}{|c|}{$\chi_l = 15$\quad$x_l=4$\quad $x_d=(3\ 3 \ 3)$
\quad $x_q=x_u=(4\ 2\ 0)$ \quad$x_e= (-1 \, 1\, 3)$}\\ 
\hline \hline 
$-6$\quad$+3$\quad$\ 0$\quad$-3$\quad$+4$ & $Q$ or $E$&$m_Q 
\sim m_E \sim \epsilon^{10} \,\Lambda$& $Q \to qql$ & $\sim 10^{9}$ km \\ 
$+4$\quad $-5$\quad$+1$\quad$-4$\quad$-6$  & & $\sim 1$ TeV
& $E \to qqu $ & $\sim 10^{9}$ km \\ 
\hline \hline 
$-6$\quad$+3$\quad$+5$\quad$+5$\quad $+6$  & $E$ & $m_E \sim
\epsilon^{11} \,\Lambda$& $E \to qqu$ & $\sim 5 \times 10^{12}$ km  \\ 
$+4$\quad $-4$\quad$+5$\quad $+4$\quad$-5$ & & $\sim 0.2$ TeV & & \\  
\hline \hline 
$-1$\quad$+2$\quad$-3$\quad$-6$\quad$+3$  & $U$ & $m_U \sim 
\epsilon^{10} \,\Lambda$& $U \to ued,uql$ & $\sim 10^{9}$ km \\ 
$+7$\quad$-8$\quad$-4$\quad$-1$\quad$-1$ & &$\sim 1$ TeV & &\\ 
\hline
\multicolumn{5}{|c|}{$\chi_l = -15$\quad$x_l=-4$\quad$x_e= (0 \, -2\, -4)$ 
\quad $x_q = x_u + 2 = (5 \ 3\ 1)$\quad $x_d= (2\  2\  2)$}\\ 
\hline \hline 
$-13$~~$+4$\quad$-2$\quad$-6$\quad$+3$ & $E$ & $m_E \sim  
\epsilon^{9} \,\Lambda$&  $E \to qqu$ & $\sim 10^{7}$ km \\ 
$-6$\quad$-2$\quad$-9$~~$-13$~~$+12$  & & $\sim 0.3$ TeV
&    &   \\ 
\hline
\end{tabular}
\caption{\label{tab3}Examples of solutions to the anomaly cancellation
  equations, with $x = X-Z'$, for some choices of the charges in the
  light sector. In each case, the lightest metastable particle (LLLP),
  and its mass, decay modes and mean path are displayed.}
\end{table}

\section{Experimental searches: production and detection}

\begin{figure}[t]
\includegraphics[width=2.75in]{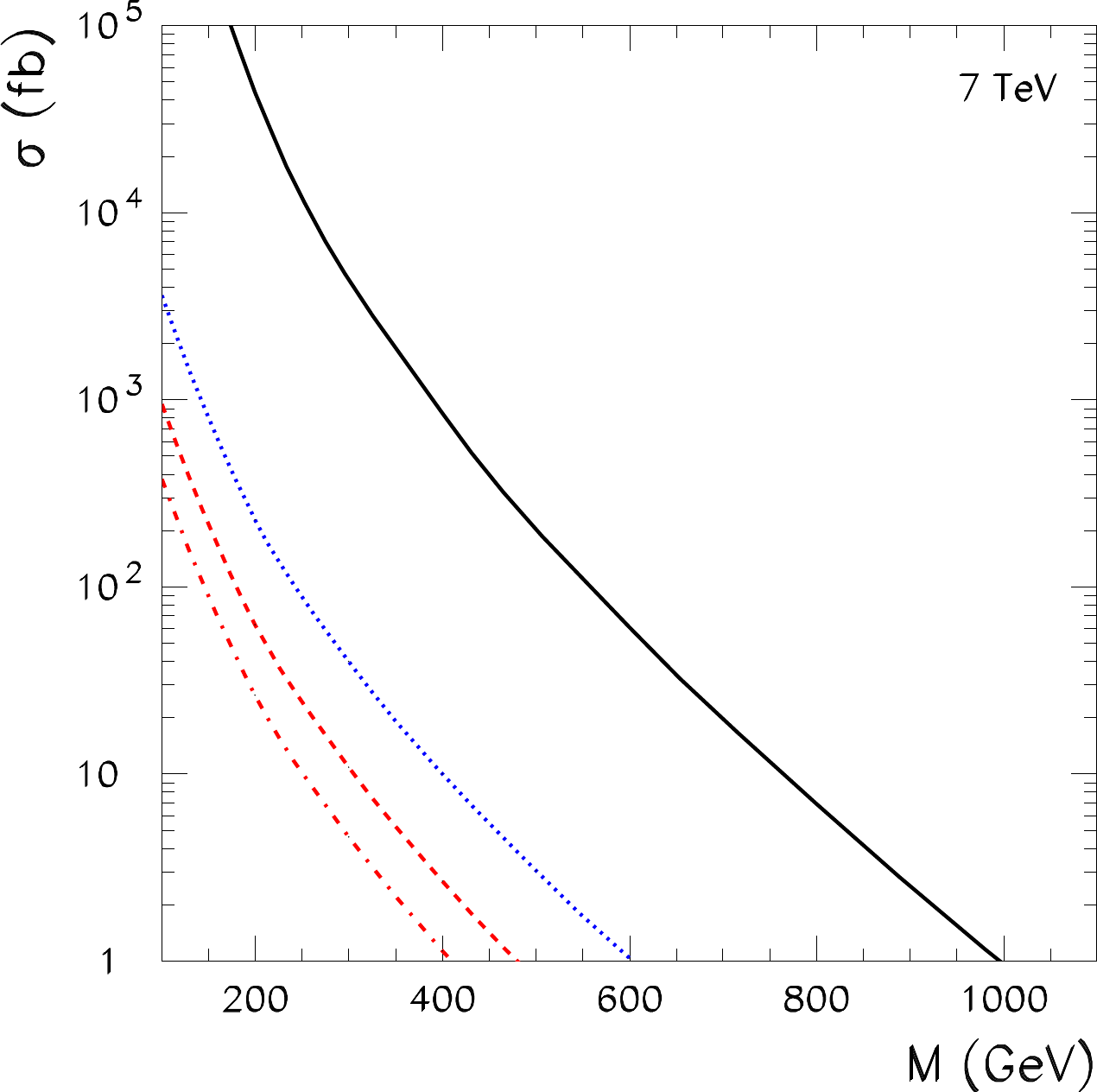}
\hfill
\includegraphics[width=2.95in]{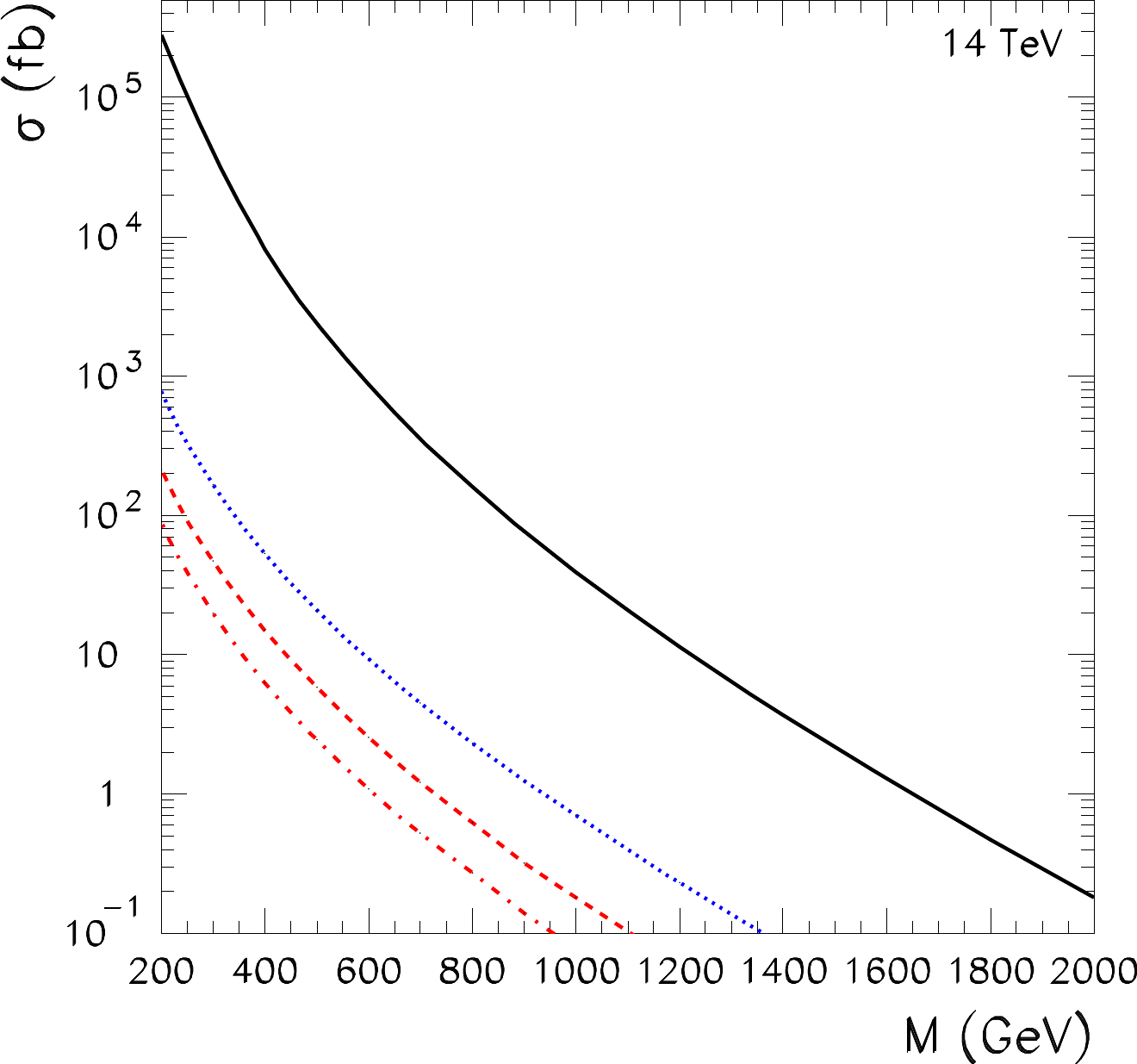}
\caption{ Total production cross section at the LHC as a function of
  the mass of the LLLP at a center-of-mass energy of 7 TeV (left
  panel) and 14 TeV (right panel). The black solid line stands for the
  strong interaction production of new heavy quarks while the blue
  dotted, red dashed and red dot-dashed lines represent the production
  of doublet charged lepton pairs, doublet charged and neutral lepton
  pairs and singlet charged leptons, respectively.}
\label{fig:sig}
\end{figure}

\noindent The new heavy quarks and leptons can be pair produced at the
LHC with sizeable cross sections as shown in figure~\ref{fig:sig} for
center--of--mass energies of 7 and 14 TeV. As seen above, the decay of
the new heavy states takes place through suppressed operators, leading
to long lifetimes.  The search for new long-lived states is conducted
by the ATLAS and CMS collaboration using several experimental
techniques, such as $dE/dx$, time of flight and decays during beam
collision intervals~\cite{exp}. 

The out--of--time decays~\cite{exp2} constrains the heavy particle
production cross section times stopping probability to be smaller than
$\simeq 0.2$ fb. Assuming that the stopping probability for the hadron
containing new heavy quarks is similar to the one for stop
hadrons~\cite{exp}, i.e., of the order of 20\%, the presently
available data exclude new long--lived quarks with mass $\lesssim 400$
GeV. Certainly, this bound should be taken with a pitch of salt since
it depends upon the unknown strong interactions of hadrons exhibiting
a heavy quark. In the case of new heavy leptons, the out--of--time
analysis excludes new leptons with masses $\lesssim 100$ GeV, assuming
that the stopping probability is equal to the one for staus,  i.e.
5\%. 

More stringent limits can be obtained from the search for slowly
moving charged particles~\cite{exp2}. Assuming that the detection of
hadrons containing heavy quarks is similar to the one for hadrons
possessing stops we obtain that the new heavy quarks must have a mass
in excess of $\simeq 800$ GeV.  Analogously, the search for slow moving
staus can be translated to an lower limit on the new lepton mass of
$\simeq 400$ GeV. \smallskip

\section{Conclusions}

We have presented a new class of quasi--stable hadrons and leptons
that naturally arise in the framework of flavour models. Those
particles can plausibly be expected to show up at the LHC as their
masses are protected by the flavour symmetry and can lie in the TeV
range. This is much lower than the symmetry breaking scale that is
constrained by FCNC and CP violating experimental data to be $\Lambda
> 5 \cdot 10^{4}$ TeV.  Limits on the new heavy fermions from EWPT can
be avoided by suppressing mixed couplings of the new fermions with the
SM ones through the Higgs by the same residual proton--hexality
symmetry that prevents rapid proton decay and explains the observed
texture of fermion masses and mixings.  Presently there are ongoing
searches at the LHC that can reveal their existence.


\acknowledgments{The work of OJPE and RZF was partially supported by
  Conselho Nacional de Ci\^encia e Tecnologia (CNPq).  The work of CAS
  was partially supported by the European Commission under the
  contracts PITN-GA-2009-237920.  OJPE and RZF thank the Institut de
  Physique Th\'eorique de Saclay for their hospitality during our
  visit where part of this work was done.}


\end{document}